\documentclass[conference]{IEEEtran}
\IEEEoverridecommandlockouts
\usepackage{cite}
\usepackage{amsmath,amssymb,amsfonts}
\usepackage[noend]{algpseudocode}
\usepackage{algorithm}
\usepackage{graphicx}
\usepackage{textcomp}
\usepackage{xcolor}
\usepackage{verbatim}
\usepackage{url}
\usepackage{array}
\usepackage{amssymb}

\def\BibTeX{{\rm B\kern-.05em{\sc i\kern-.025em b}\kern-.08em
    T\kern-.1667em\lower.7ex\hbox{E}\kern-.125emX}}
    \makeatletter
\newcommand{\linebreakand}{%
  \end{@IEEEauthorhalign}
  \hfill\mbox{}\par
  \mbox{}\hfill\begin{@IEEEauthorhalign}
}
\makeatother

\begin{document}

\title{Automated Gateways: A Smart Contract-Powered Solution for Interoperability Across Blockchains}

\author{\IEEEauthorblockN{1\textsuperscript{st} Koosha Esmaeilzadeh Khorasani}
\IEEEauthorblockA{\textit{Department of Computer Science} \\
\textit{University of Manitoba}\\
Winnipeg, Canada \\
esmaeilk@myumanitoba.ca}
\and
\IEEEauthorblockN{2\textsuperscript{nd} Sara Rouhani}
\IEEEauthorblockA{\textit{Department of Computer Science} \\
\textit{University of Manitoba}\\
Winnipeg, Canada \\
sara.rouhani@umanitoba.ca}
\and
\IEEEauthorblockN{3\textsuperscript{rd} Rui Pan}
\IEEEauthorblockA{\textit{Department of Engineering} \\
\textit{Grain Discovery}\\
Edmonton, Canada \\
rui@graindiscovery.com}
\linebreakand

\IEEEauthorblockN{4\textsuperscript{th}  Vahid Pourheidari}
 \IEEEauthorblockA{
 \textit{Department of Engineering}\\
\textit{Futurix Technologies}\\
Winnipeg, Canada \\
vahid.p@futurix.ca}
}

\maketitle

\begin{abstract}
Interoperability is a significant challenge in blockchain technology, hindering seamless data and service sharing across diverse blockchain networks. This study introduces \textit {Automated Gateways} as a novel framework leveraging smart contracts to facilitate interoperability. Unlike existing solutions, which often require adopting new technologies or relying on external services, Automated Gateways framework is integrated directly with a blockchain's core infrastructure to enhance systems with built-in interoperability features. By implementing fine-grained access control mechanisms, smart contracts within this framework manage accessibility and authorization for cross-chain interactions and facilitate streamlining the selective sharing of services between blockchains. Our evaluation demonstrates the framework's capability to handle cross-chain interactions efficiently, significantly reduce operational complexities, and uphold transactional integrity and security across different blockchain networks. With its focus on user-friendliness, self-managed permissions, and independence from external platforms, this framework is designed to achieve broader adoption within the blockchain community. 

\end{abstract}


\section{Introduction}
Blockchain technology is increasingly recognized for its potential to enhance various industries \cite{reegu2021interoperability,gordon2018blockchain, wang2019loc,wang2019blockchain, sajja2023towards,iranmanesh2023effects}, by transitioning from traditional centralized server-based networks to decentralized, transparent, and secure systems and marking a significant shift in how data and transactions are managed and verified across various sectors.

The growing interest and investment in blockchain technology have led to the development of various blockchain platforms, each with its unique features and advancements. However, a common challenge with these platforms is the lack of built-in interoperability. They are often designed to work in isolation, without the capability to communicate or integrate services directly with other blockchain systems. This isolation results in data and assets being trapped within individual blockchains, creating silos that compromise the efficiency \cite{karumba2023bailif, wang2021sok} and trustworthiness \cite{abebe2019enabling, schulte2019towards} of the blockchain ecosystem.
Interoperability in the blockchain context is the capability for diverse blockchain systems to collaborate seamlessly which is crucial for executing transactions across multiple blockchain platforms and ensuring the acknowledgment, access, and verification of data recorded on one blockchain by another. Interoperability among blockchains is essential for creating a unified infrastructure, fostering innovation, and expanding the functionality of decentralized applications across multiple platforms. \cite{yaga2019blockchain, wang2023exploring}. 

Although various interoperability solutions such as Polkadot\cite{wood2016polkadot}, Quant Overledger \cite{verdian2018quant}, Axelar\cite{Axelar}, WanChain\cite{wanchain2017} and Cosmos\cite{kwon2019cosmos} have been developed, their lack of widespread adoption still leaves data and service silos as an unresolved issue. One major barrier to the practical application of these solutions is the extensive time and effort required for development teams to integrate them into existing platforms. This integration process involves a steep learning curve because developers must understand each platform's unique architecture, APIs, and design principles. For some solutions, there is also a need to construct and maintain specialized bridges \cite{xie2022zkbridge}, leading to extra development and maintenance costs. Moreover, relying on external support risks service discontinuation, complicating interoperability adoption. Another issue is that existing solutions mainly focus on public blockchains and ignore authentication and access management in permissioned blockchains, a gap in the literature.

We propose \textit {Automated Gateways} framework, a novel blockchain interoperability solution designed to address challenges in data and service sharing across different blockchain platforms. Our solution relies on existing blockchain platforms, minimizing the need for external interoperability services. By using smart contracts within blockchain platforms to implement fine-grained access control mechanisms, authorization policies, and selective sharing of specific services between blockchains, it minimizes the need for new technologies and reduces reliance on third-party services. Additionally, the framework features a user-centric design, with intuitive APIs that streamline blockchain interconnectivity with minimal coding.

To implement and evaluate our solution, we developed a distributed production network instead of a local test network. This method transcends the limitations of local simulation and test networks typically employed in similar studies. \cite{scheid2019bifrost, rahman2022blockchain}. 

The remainder of this paper is organized as follows. Section \ref{related_works} provides a summary of related studies on blockchain interoperability and existing frameworks. Section \ref{system_design} presents the architecture, security overview, and process flow of the proposed framework. Section \ref{impl_eval} discusses the implementation and evaluation of the Automated Gateways framework. Finally, Section \ref{conclusion} draws conclusions from our findings and outlines the potential future directions for research and development.

\section{Related Works}
\label{related_works}
The literature on blockchain interoperability is extensive, including numerous methods across diverse architectures and use cases. Numerous survey papers offer different frameworks for understanding and categorizing blockchain interoperability. They investigated aspects such as interoperability layers  \cite{lohachab2021towards}, architecture \cite{belchior2021survey,kannengiesser2020bridges,wang2023exploring}, the stability of interoperability solutions \cite{zuo2021review}, and the integration of smart contracts as a critical component in achieving interoperability\cite{khan2021towards}. Security concerns, a fundamental aspect of blockchain interoperability, are comprehensively examined in several surveys, which discuss various strategies to secure blockchain communications \cite{haugum2022security,zhang2023sok,lee2023sok,augusto2023sok}. 

Blockchain interoperability solutions are generally divided into five categories: Sidechains, Trusted Relays, Notary Schemes, Hashed Time-Lock Contracts (HTLC), and Blockchain of Blockchains \cite{ren2023interoperability, belchior2021survey}.

\textbf{Sidechain}  is an independent blockchain operating alongside a mainchain, allowing secure asset and data exchanges through a cross-chain communication protocol, achieving interoperability with a two-way peg mechanism\cite{back2014enabling}.
.
An instance of a sidechain-based solution is The Loom Network \cite{loomxio}, which employs a federated two-way peg scheme for asset swapping. Another sidechain-based platform is RootStock (RSK)\cite{lerner2019}, which includes a Turing-complete virtual machine for smart contracts, a two-way pegged Bitcoin sidechain backed by a federation with custom Hardware Security Modules (HSM-modules), a consensus protocol resistant to selfish mining, and a low-latency block-propagation network. Another example is Liquid\cite{liquid2020}, which facilitates the transfer of bitcoins into Liquid Network Assets using a cryptographic peg based on Sidechain architecture.

\textbf{Trusted Relays} act as direct links between blockchains, bypassing the need for intermediaries. In this scheme, blockchain A incorporates a smart contract to enable verification of data on blockchain B. The smart contract retrieves the block header of a specific block on blockchain B and verifies it according to the procedure specified by blockchain A's consensus algorithm \cite{buterin2016chain}. An example of Trusted Relay architecture is  ETH Relay \cite{Eth_relay}, aiming to reduce the cost associated with Simplified Payment Verification (SPV) \cite{nakamoto2008bitcoin} by employing an on-demand verification approach. Another example is Weaver \cite{weaver_dltinteroperability} in which the architectural structure is similar to \cite{abebe2019enabling}. Weaver includes relays for communication, and its interoperability (IOP) modules encapsulate the logic governing membership, verification policies, and access control policies. 

\textbf{Notary Scheme} solution is characterized as a trusted witness or group of witnesses tasked with validating the claims of typically untrusted blockchains involved in interoperability-based processes, such as proof of asset ownership \cite{wang2023exploring}.
An illustration of a Notary Scheme platform is Bifröst \cite{scheid2019bifrost}. Bifröst oversees the management of public and private keys and the credentials hash of transactions. BAILIF \cite{karumba2023bailif}  is an interoperability  solution integrates relays and a Decentralized Notary Service (DNS). Notary nodes are selected based on the DNS algorithm, which considers reputation, reliability, and availability. These nodes verify transaction validity, generate, and validate the Merkel proof.

\textbf{Hashed Time-Lock Contracts (HTLC)} protocol relies on a combination of hash-lock \cite{hashlock} and time-lock \cite{timelock} mechanisms. To facilitate an atomic cross-chain transaction in HTLC solutions, the transaction source must furnish cryptographic proof before a predetermined deadline \cite{belchior2021survey}. 
Black, Liu, and Cai \cite{black2019atomic} employed the HTLC mechanism to implement a blockchain-based loaning system. The HTLC mechanism is also utilized in Bitcoin Lightning Network \cite{poon2016bitcoin}, which serves as a payment channel network designed to address scalability challenges within the Bitcoin platform by establishing multiple hops of payment channels to support cross-blockchain atomic swaps.

\textbf{Blockchain of Blockchains (BoB)} is an interoperability solution wherein a blockchain is designated to store and track transactions generated by DApps distributed across multiple blockchains \cite{belchior2021survey}.
 Polkadot \cite{wood2016polkadot} and Cosmos \cite{kwon2019cosmos} are well-known examples of BoB architecture, which achieve interoperability through their main blockchain and parallel chains, which are interconnected using the main blockchain. Another solution, Axelar \cite{team2023axelar}, similarly employs a central blockchain to facilitate interoperability using validators and relayers. 

 Automated Gateways solution proposed by this study presents distinct advantages in integration, security, and operational efficiency compared to traditional blockchain interoperability methods, as detailed in the provided comparison table \ref{litrature-comparision}. Unlike sidechains that require a secondary blockchain or trusted relays that necessitate specific infrastructure additions, Automated Gateways utilize direct smart contract integration to enhance security and reduce dependencies, making them inherently less prone to trust issues prevalent in notary schemes and trusted relays. Furthermore, Automated Gateways exhibit superior efficiency and decentralization characteristics compared to HTLC and Blockchain of Blockchains (BoB) solutions, which can suffer from conditional efficiency and semi-centralized governance.

\begin{table*}[!ht]
\centering
\caption{Comparison of Blockchain Interoperability Solutions. Symbols used: $\checkmark$ indicates a positive feature or low dependency, $\times$ indicates a negative feature or high dependency, and $\bigtriangleup$ indicates moderate characteristics or dependency.}
\label{litrature-comparision}
\footnotesize
\begin{tabular}{|p{2cm}|p{2.3cm}|p{2.3cm}|p{2.3cm}|p{2.3cm}|p{2.3cm}|p{2.1cm}|}
\hline
\textbf{Feature} & \textbf{Automated Gateways} & \textbf{Sidechains} & \textbf{Trusted Relays} & \textbf{Notary Schemes} & \textbf{HTLC} & \textbf{BoB} \\ \hline
\textbf{Integration} & Direct with SC & Secondary blockchain required & Specific relay infrastructure & Trusted entities for validation & Direct, condition-based & New main blockchain \\ \hline
\textbf{Dependency} & Low ($\checkmark$) & Moderate ($\bigtriangleup$) & High ($\times$) & High ($\times$) & Low ($\checkmark$) & High ($\times$) \\ \hline
\textbf{Security} & Fine-grained SC controls & Entities involved in pegging mechanism & Security mechanism of the relay component &  Monitoring trusted entities & Cryptographic proofs & Security of the middle blockchain \\ \hline
\textbf{Efficiency} & High ($\checkmark$) & Operational limits ($\bigtriangleup$) & Adds complexity ($\times$) & Slow validation process ($\times$) & Conditional, efficient ($\bigtriangleup$) & Infrastructure heavy ($\times$) \\ \hline
\textbf{Decentralization} & High ($\checkmark$) & Under main chain influence ($\bigtriangleup$) & Semi-decentralized ($\bigtriangleup$) & Decentralized ($\checkmark$) & Decentralized ($\checkmark$) & Semi-centralized ($\times$) \\ \hline
\textbf{Transparency} & High ($\checkmark$) & Within ecosystem ($\bigtriangleup$) & Relay dependent ($\bigtriangleup$) & Trust dependent ($\times$) & Transparent ($\checkmark$) & Varies ($\bigtriangleup$) \\ \hline
\end{tabular}
\end{table*}

\section{System Design}
\label{system_design}
This section provides an overview of the key design principles used in developing Automated Gateways, along with the architecture and integral components of the proposed solution. It also presents a detailed security overview and process flow to explain the solution's functioning.

\subsection{Design Principals}
The Automated Gateways solution is grounded in three fundamental principles, serving as its cornerstone:
\begin{enumerate}
  \item \textbf{Decentralization:} Embracing the core tenet of Web3, our proposed solution prioritizes decentralization to mitigate the risks associated with a single point of failure and the reliance on trusted third parties.

  \item \textbf{Transparency:}Transparency is crucial for sharing services and data, as it supports trust-building, auditability, accountability, and verifiability. Our interoperability solution incorporates transparency by using smart contracts and on-chain transactions to manage policies.
  \item \textbf{Self-reliance:} 
  Our proposed smart contract-based solution emphasizes self-reliance by using existing technologies in the host blockchain platform, minimizing the need for new technologies, reducing reliance on third-party solutions, and minimizing changes to the platform's architecture.

\end{enumerate}
\subsection{Architecture}
\begin{figure*}[!tbp]
\includegraphics[width=\textwidth ,height=6cm]{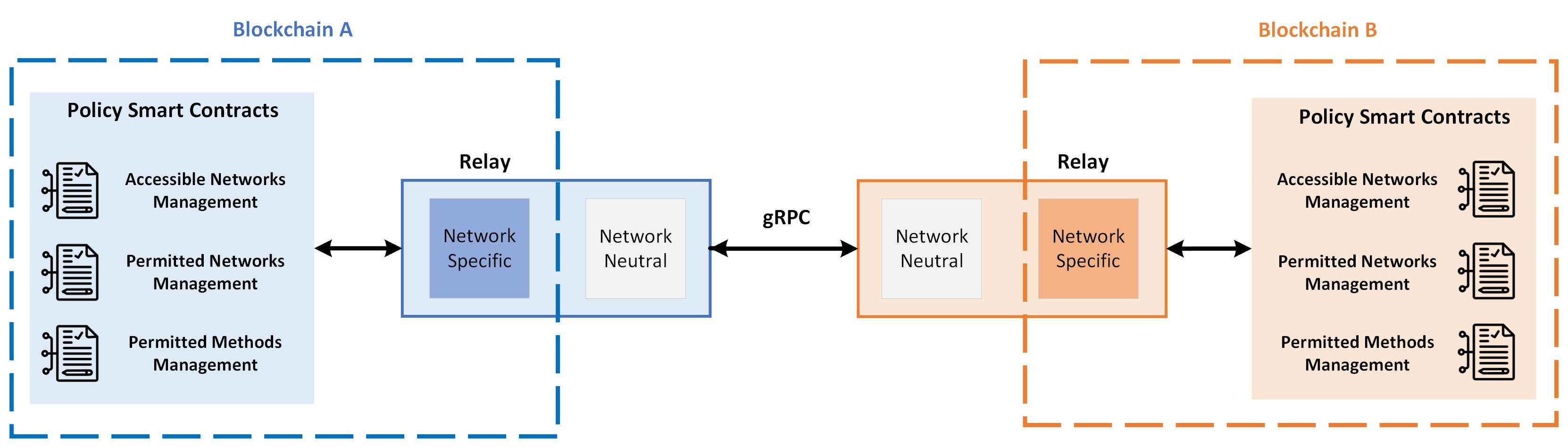}
\caption{The high-level architecture diagram of Automated Gateways.}
\label{architecture}
\end{figure*}
To align with our design principles, the proposed solution is organized into two key components: Policy Management and Communication Management. As illustrated in Figure \ref{architecture}, the Policy Management component operates through a set of smart contracts, referred to as 'policy smart contracts'. They are deployed on the host blockchain to ensure distribution, transparency, and self-reliance. Distribution is ensured by deploying smart contracts across various nodes in the blockchain network. Transparency is achieved through transactions recorded on a shared ledger, and self-reliance is maintained by leveraging existing blockchain systems, rather than developing external services.

The Communication Management component, referred to as the relay, serves as a messenger for transferring data and services between different blockchains using the gRPC\cite{indrasiri2020grpc} protocol. To achieve distribution within a blockchain, a relay must be deployed for each peer-hosting policy smart contract. The relay was developed and released as a library to streamline the development and enhance adaptability. This allows easy integration into any source code used by blockchain admins, thereby minimizing the time and effort required for implementation. A relay components is divided into two parts. The first part is network-neutral, meaning it is blockchain-agnostic and designed to operate independently of the host blockchain. It manages communication between blockchains and executes user commands. The second part is network-specific, responsible for connecting the relay to the host blockchain. This part utilizes the host blockchain's software development kit (SDK) to establish and manage the connection. The following sections provide a detailed exploration of the Policy Management and Communication Management components.

\subsubsection{Policy Management Component}
This section comprises three smart contracts designed to establish authentication and authorization policies. These three smart contracts include the following:
\begin{itemize}
    
  \item \textbf{Accessible Networks Smart Contract:} This smart contract manages the blockchain networks accessible to the host blockchain, using its stored data to establish connections with these networks.
  \item \textbf{Permitted  Networks Smart Contract:} The Permitted Networks Smart Contract is responsible for managing information about blockchains that have been granted access to the host blockchain data or services, allowing them to proceed with requests for desired services and data on the host blockchain.
  \item \textbf{Permitted Methods Smart Contract:}  The Permitted Methods Smart Contract introduces an additional layer of authorization to our system. While the Permitted Networks Smart Contract grants authorization to networks accessing data and services on the host blockchain, the Permitted Methods Smart Contract specifies the precise services and methods within the host blockchain that other networks can invoke and access. 
\end{itemize}

\subsubsection{Communication Management Component}

\begin{figure}
\includegraphics[width=0.5\textwidth ,height=6cm]{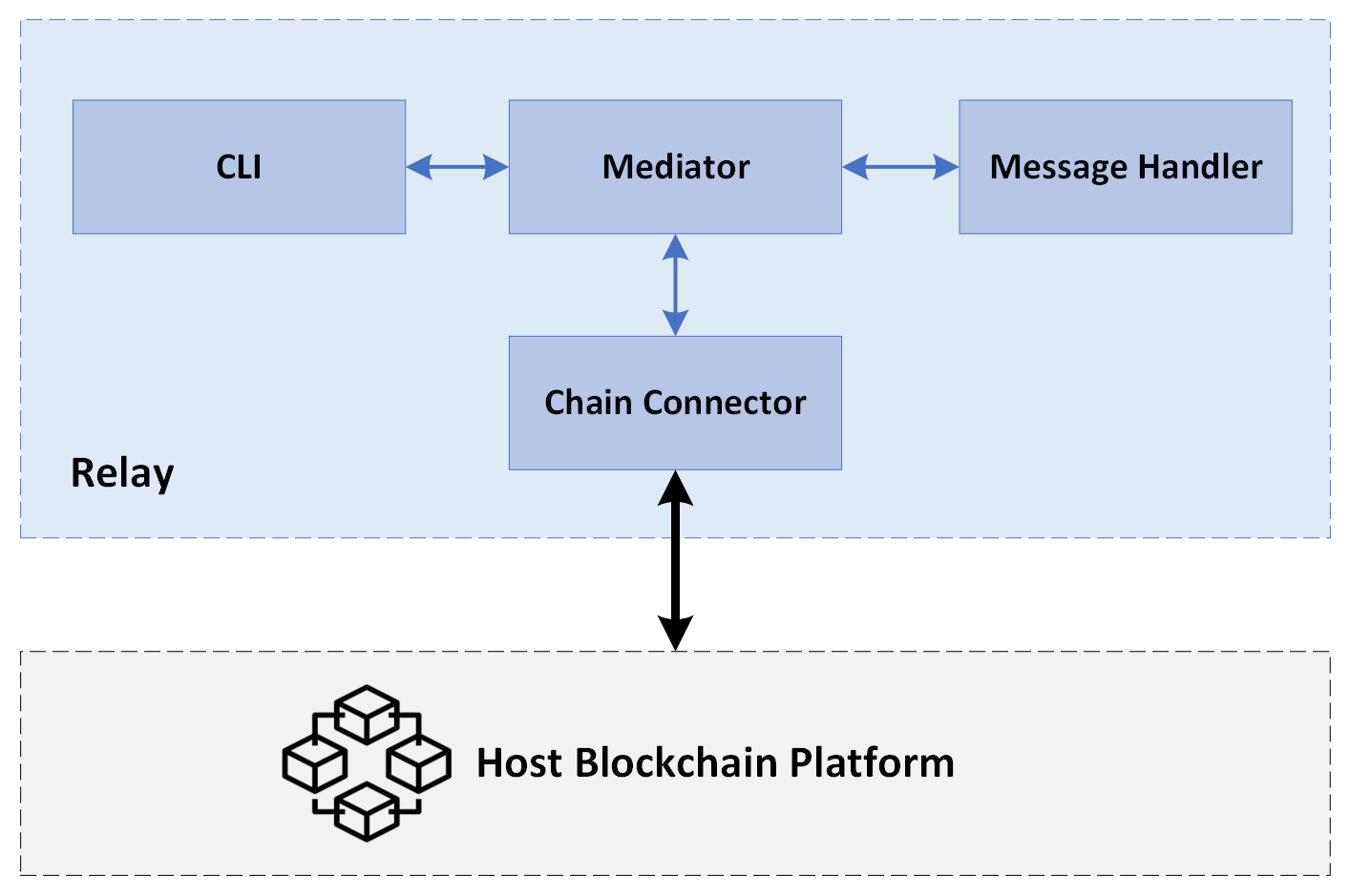}
\caption{Overview of the Relay Component.}
\label{relay}
\end{figure}
The Communication component is pivotal for establishing connections between the host and permitted blockchains, enabling the exchange of data and services. As depicted in Figure \ref{relay}, this entity includes four main components: a Chain Connector, which represents the network-specific part of the relay, and a Mediator, Message Handler, and Command Line Interface (CLI), which constitute the network-neutral part of the relay.

The \textbf{Chain Connector} is the first component, responsible for linking the relay with the host blockchain. It manages all interactions between them, such as interactions with policy smart contracts, data retrieval, and service invocation. Different Chain Connectors are needed to support various types of host blockchain networks, with each implementation based on the software development kit provided by the creators of the respective host blockchain.

The \textbf{Mediator} is the second component of the relay, primarily responsible for connecting different elements within the relay. All requests for sending or receiving data or services are routed through the Mediator, and corresponding responses are handled by it. Additionally, this component provides APIs to users, enabling them to access relay services and initiate the interoperability process to send or receive data from the host or accessible blockchains.

The \textbf{Message Handler} component acts as the central hub for communication between the host blockchain and accessible blockchains. We chose gRPC as our connection protocol due to its advantages, including a simple and well-defined interface, seamless integration with the cloud ecosystem, and support for multiple programming languages. 

The final component is the \textbf{Command Line Interface (CLI)}, designed to offer system admins, who are in charge of configuring and managing the host blockchain, a user-friendly way to interact with the interoperability platform without requiring coding skills. Through the CLI, admins can manage information about shared services and data, oversee accessible and permitted networks, and administer authorization policies. Moreover, they can invoke services that are accessible to the host blockchain.
\subsection{Security Overview}
To enable the secure exchange of data and services, it is necessary to design an approach that ensures the security and confidentiality of interactions, both between two relays and between the relay and the host blockchain.

To establish a secure flow of data between two relays, we implemented certificate-based authentication\cite{lal2016review}. Certificate-based authentication is a process wherein a digital certificate is employed to verify the identity of a user, device, or system. In this context, the admin of the host blockchain generates certificates for its relay server and issues certificates for the relay clients of blockchains seeking to connect to the host blockchain (permitted blockchain). The host blockchain admin can utilize its own certificate authority or any valid certificate authority of its choice. After registering an external blockchain as a permitted blockchain, the host blockchain admin should provide the permitted blockchain admin with a client certificate. The permitted blockchain admin can then introduce this certificate to the relay through relay configuration. This certificate plays a crucial role in the authentication process, as described in Algorithm 1, which is applied to all the requests received by the gRPC server.

\begin{algorithm}
\caption{Certificate authentication algorithm}\label{alg:one}
\begin{algorithmic}[1]
\Procedure{Authentication}{$CertInfo$}
\State $cn\gets  extarctCommonName(CertInfo)$
\State $pn \gets GetPermittedNetworks(cn)$\Comment{get from chain}
\If{$pn$ exists}
\State \textbf{return} true \Comment{authentication succeeded}
\Else
\State \textbf{return} false \Comment{authentication failed}
\EndIf
\EndProcedure
\end{algorithmic}
\end{algorithm}
As depicted in Algorithm \ref{alg:one}, the authentication process between two relays initiates by extracting the common name, denoting the URL for which the certificate is issued, from the certificate data of the relay of a permitted blockchain. Then, a verification, in the host blockchain, is conducted against the Permitted Networks Management Contract to check whether this URL is registered. If the URL is registered as a permitted network, the request proceeds for further processing; otherwise, an error is returned.

To ensure secure data transfer between the relay and the host blockchain, the relay adheres to the protocol established by the respective blockchain network. As an illustration, Hyperledger Fabric employs certificate-based authentication \cite{hlfapp}, and our connection security aligns with this established protocol to connect the relay to the Hyperledger Fabric blockchain network.

\subsection{Process Flow}
The process of data and service sharing in Automated Gateways comprises two distinct steps. The first step involves registration, during which the necessary configurations are established to enable desired access between blockchains. The second step involves invoking methods that enable blockchains to share data and services. In order to discuss the designed flow, we considered a scenario with two blockchains, Blockchain(1) and Blockchain(2), where we aim to outline a flow allowing Blockchain(2) to access a service provided by Blockchain(1).

To successfully complete the registration process, the following steps must be accomplished:

\begin{enumerate}
\item \textbf{Request for Access:} This step initiates outside the domain of computer systems, where the admins of Blockchain(2) engages in negotiations with the admins of Blockchain(1) to obtain access to a specific service provided by one of the smart contracts on Blockchain(1).
\item \textbf{Certificate Issuance:} In case of getting permission, the admins of Blockchain(1) must issue certificates to the admins of Blockchain(2) specifically for the purpose of certificate-based authentication.
\item \textbf{On-chain Registration:} Following the issuance of certificates, the admins of Blockchain(1) are required to register Blockchain(2) as a permitted network, while the admins of Blockchain(2) should reciprocate by registering Blockchain(1) as an accessible network. Moreover, To enable access to the desired service, the admins of Blockchain(1) must grant permission to Blockchain(2). This is done by registering the method information and the Permitted Network ID of Blockchain(2) as a permitted method entity in the Permitted Methods Smart Contract.
\end{enumerate}

\begin{figure*}
\centering
\includegraphics[width=0.8\textwidth ,height=8cm]{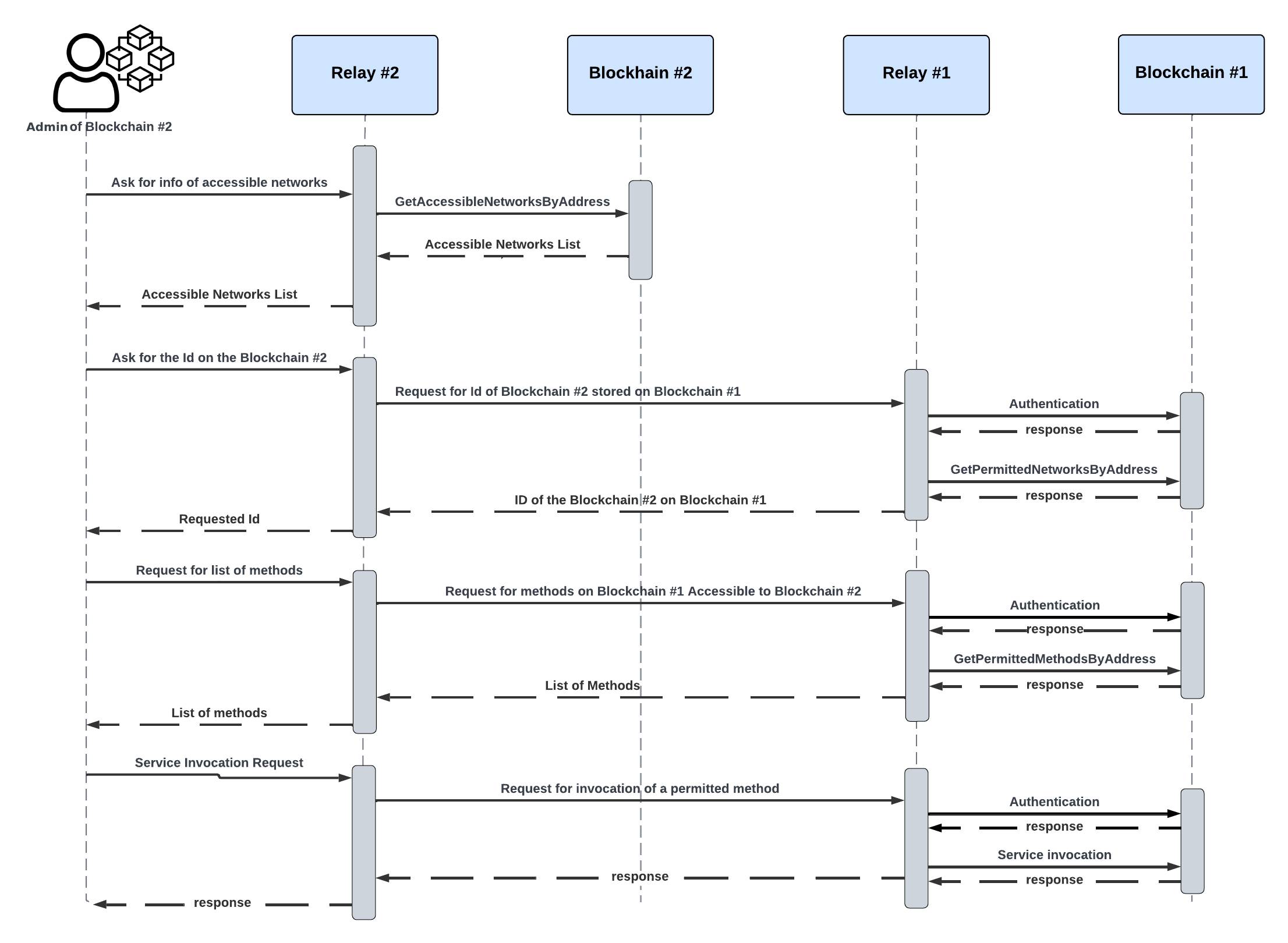}
\caption{The sequence diagram of cross-platform service invocation.}
\label{invocationStep}
\end{figure*}  

In regard to the method invocation process, as it is illustrated in Figure \ref{invocationStep}, the admin of the blockchain(2) should follow these steps:
\begin{enumerate}
\item \textbf{Get Accessible Network Information:} For any communication between two blockchain relays, a connection must be established through the Message Handler component. To configure the Message Handler component of the relay, specific information about the destination accessible network, including its IP address, needs to be obtained from the blockchain. The admin of Blockchain(2) obtains information about the desired accessible network by interacting with the relay through the Command Line Interface (CLI). The CLI receives the admin's request and transmits it to the Mediator, which in turn sends it to the Chain Connector. Chain Connector gets the response by calling GetAccessibleNetworksByAddress from the Accessible Networks Smart Contract. The response from the Blockchain(2) follows the reverse stream back to the admin.
\item \textbf{Request for Permitted Network Information:} In this step, Blockchain(2) sends a request to Blockchain(1) to obtain its unique identifier stored on Blockchain(1). This identifier will be used in the next step to retrieve the list of methods accessible to Blockchain(2). When the relay of Blockchain(1) receives the request for the desired information, it verifies the certificate information of Blockchain(2) relay with Blockchain(1) to authenticate the process, following the steps outlined in algorithm \ref{alg:one}. Upon successful authentication, Blockchain(1) relay, which we call relay(1), retrieves information from Blockchain(1) using the GetPermittedNetworksByAddress method from the Permitted Networks Smart Contract and shares it with Blockchain(2) relay, which we call relay(2).
\item \textbf{Request for Permitted Methods Information:}  To utilize a service or obtain data, the admin of Blockchain(2) needs information about the method on Blockchain(1) through which this data or service is accessible. To achieve this, the admin of Blockchain(2) sends a request to relay(1) to obtain a list of methods using relays(2). After successful authentication, relay(1) acquires the list of permitted methods for Blockchain(2) using its unique identifier. The Chain Connector of relay(1) invokes the GetPermittedMethodsByNetworkId method from the Permitted Methods Smart Contract. Upon receiving the response from the chain, relay(1) forwards the response to relay(2). Subsequently, the admin of Blockchain(2) can retrieve the list of available methods.
\item \textbf{Service Invocation:} After obtaining the list of methods and their information, the admin of Blockchain(2) can request the invocation of specific methods to access the provided data or service. relay(1) forwards the information to the Chain Connector via the Mediator component after authentication. The Chain Connector then invokes the desired method and returns the result to the Mediator. The Mediator forwards it to the Message Handler, which sends the response back to relay(2). This enables the admin of Blockchain(2) to retrieve and utilize the data or service result.
\end{enumerate}

\section{Implementation and Evaluation}
\label{impl_eval}
In this section, we detail the implementation of the proposed architecture, including the development technologies and deployment infrastructure. We also present the technologies used to evaluate the Automated Gateways framework and the results of the performance analysis. 

\subsection{Implementation}
For the development of Automated Gateways framework, Go programming language was chosen for its efficiency and suitability in creating the relay component, which was then released as a Go module. This modular design simplifies the integration process, allowing any Go program that includes Automated Gateways to initiate the relay with a few commands. Subsequently, the relay module handles the underlying processes required for interoperability.

Furthermore, Hyperledger Fabric, a mature and widely adopted permissioned blockchain platform \cite{hlfannual2020}, was selected for this study due to its customizable permission settings and industry relevance. This choice aimed to provide practical insights into the real-world applicability of our architecture. Smart contracts for Hyperledger Fabric were developed in Go, ensuring streamlined interaction between the relay and the blockchain network. Both the relay module and smart contracts are available on GitHub\footnote{\url{https://github.com/tcdt-lab/Automated-Gateways}}.

\subsection{Evaluation}
This section presents the performance evaluation results for the Automated Gateways framework, emphasizing three critical methods within the policy smart contracts that are vital to the system's workflow. Our experiments concentrated on these key methods: GetAccessibleNetworksByAddress' from the Accessible Networks Smart Contract, GetPermittedNetworksByAddress' from the Permitted Networks Smart Contract, and `GetPermittedMethodsByNetworkId' from the Permitted Methods Smart Contract.

To attain results that more accurately reflect real-world scenarios, we deployed two Hyperledger Fabric networks in production mode, and we increased the rate of concurrent requests that the gateway of a peer could handle from 500 to 2500. For this deployment, we allocated a Virtual Private Server (VPS) with 16 gigabytes of RAM, 8 vCPUs, and 500 gigabytes of storage to each instance of  Hyperledger Fabric. Within each of our Hyperledger Fabric networks, we defined three organizations—two peer organizations and one orderer organization. To simulate real-life networking scenarios between organizations, each organization was launched on a distinct Docker engine, thereby preventing connections between these organizations through Docker's local network. For benchmarking purposes, we employed Hyperledger Caliper\cite{caliper}. Hyperledger Caliper was deployed on a separate VPS with the same configuration as the Hyperledger Fabric VPS, ensuring that the resource consumption of Caliper did not impact the performance of Hyperledger Fabric.

To begin assessing the performance of the specified methods, we initiated our evaluation by adjusting the transaction input rate. During this phase, we maintained a consistent setup with 10 concurrent workers operating on Caliper, sending a total number of 30K transactions in each round of testing different transaction input rates, aiming to analyze the system's performance under varying loads. After identifying the system’s saturation point, we fixed the transaction input rate at this level and shifted our focus to modifying the number of concurrent workers as the second phase of evaluation. In each round of testing with a specific number of workers, we sent a total of 30,000 transactions. This phase of evaluation was designed to determine the most effective number of workers for optimal system performance.

\begin{figure}
\centering
\includegraphics[width=0.5\textwidth ,height=4cm]{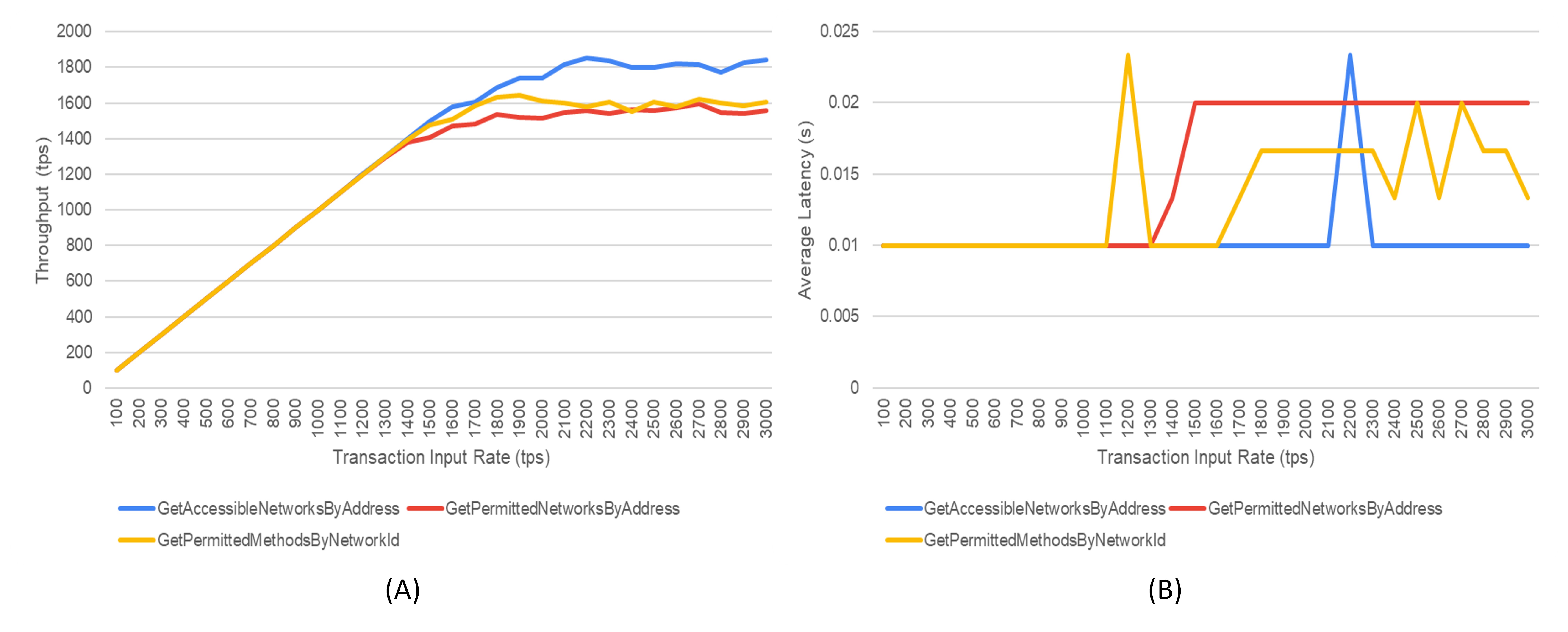}
\caption{The trends of system throughput (A) and average latency (B) across different transaction input rates.}
\label{tps_eval}
\end{figure}

As depicted in diagram (A) of Figure \ref{tps_eval}, the GetAccessibleNetworksByAddress method (blue line) indicates an initial linear correlation between input transaction rate and throughput. The throughput commenced at 100 tps and progressively increased to 1800 tps as the transaction input rate reached 2100 tps. After this point, a fluctuation will start. This trend shows that we reach the saturation point at approximately 2100 tps as the transaction input rate. This trend persisted consistently for both `GetPermittedNetworksByAddress'(red line) and `GetPermittedMethodsByNetworkId'(yellow line), with their respective saturation points identified at 1800 tps and 1700 tps, signifying a similar behavior in the performance characteristics of these methods. 

Examining the Average Latency in Figure \ref{tps_eval}, which is illustrated in diagram (B), indicates that while the latency may not consistently remain constant and some spikes may occur, particularly noticeable for the method `GetPermittedMethodsByNetworkId', the range between the maximum and minimum average latency values during various send rates does not exceed fifteen milliseconds for any of the methods, which is normal for a system with transaction input rate more than 1000 tps. This observation illustrates an acceptable level of average latency across the evaluated methods.
\begin{figure}
\centering
\includegraphics[width=0.5\textwidth ,height=4cm]{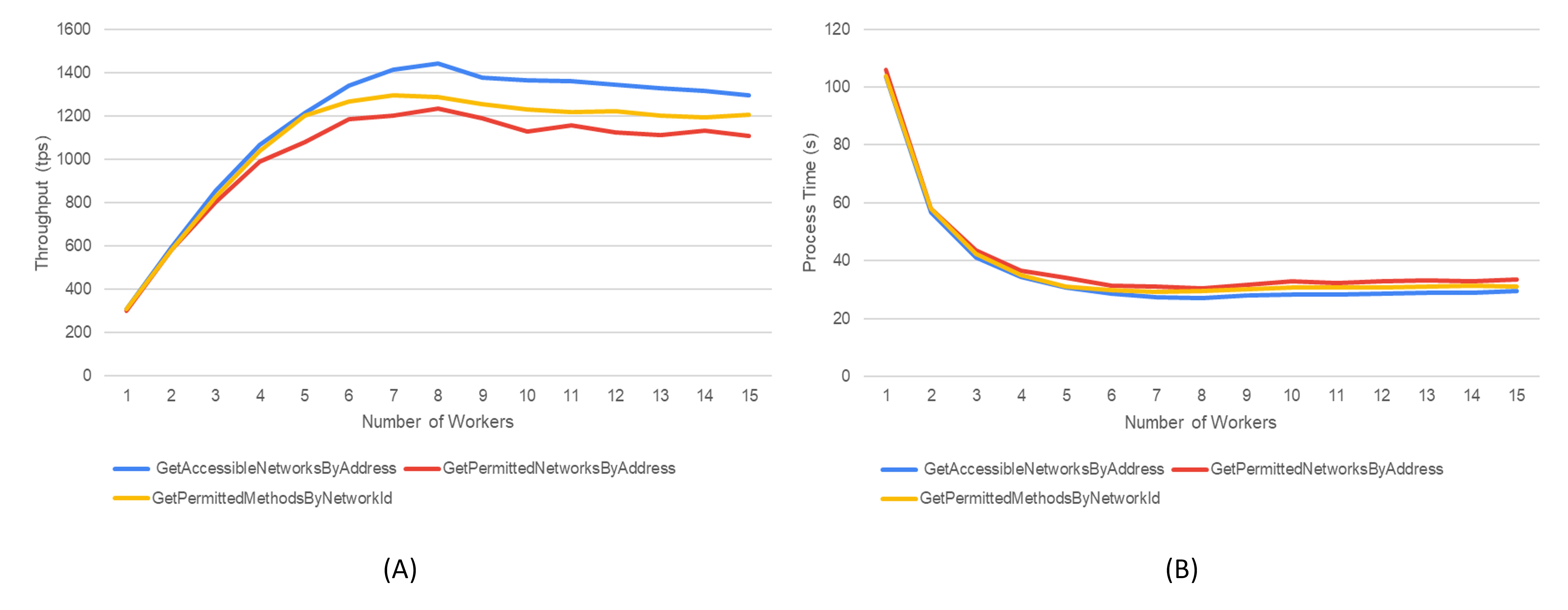}
\caption{ The trends of system throughput (A) and process time (B) across different numbers of workers.}
\label{workers_eval}
\end{figure}

In our second evaluation approach, we maintained a fixed transaction input rate at the saturation point for each method and proceeded to increase the number of workers. As depicted in diagram (A) of Figure \ref{workers_eval}, with the incremental addition of workers, an initial surge in throughput rate is observable, peaking at a certain point. Beyond this peak, the throughput plateaus. The onset of this fluctuation marks the saturation point, identified as 8 workers for `GetAccessibleNetworksByAddress', 8 workers for `GetPermittedNetworksByAddress', and 7 workers for `GetPermittedMethodsByNetworkId'.

Considering the Process Time, which is indicative of the time required for all workers to complete the test, an initial substantial decrease is noted from approximately 100 to 25 seconds before reaching the saturation point. Subsequently, a marginal fluctuation in process time is observed.

The system's ability to handle a high number of transactions per second with an average latency under 25 milliseconds demonstrates its robustness under industrial pressure. This performance confirms that our system meets our research objectives, including easy maintenance, cost-effective technology adaptation, and stable connectivity across diverse blockchain platforms.

\section{Conclusion}
\label{conclusion}
In this paper, we introduced Automated Gateways as an easy-to-use, smart contract-powered interoperability solution for facilitating data and service sharing among blockchains. The objective is to minimize reliance on external solutions and reduce the time and effort required to establish communication between blockchains. Evaluation results have confirmed the system's ability to efficiently handle a high volume of concurrent transactions with low latency, demonstrating robustness under heavy load. By deploying a production environment and carefully tuning the network and resource configurations, we achieved realistic testing conditions. Our results, which demonstrate an approximately linear correlation between transaction rate and throughput while identifying optimal worker configurations, highlight the system's scalability and efficiency. This performance aligns with our goals of easy maintenance and cost-effective integration for blockchain platforms, indicating strong potential for wider application in blockchain technology.

For the future direction, a feasible expansion of the current solution could involve integrating functionalities like atomic swap and token transfer among multiple blockchains. This extension needs to uphold the design principles and intrinsic values of Automated Gateways framework, thereby augmenting its capabilities and broadening its applicability across diverse blockchain scenarios.

\bibliographystyle{IEEEtran}
\bibliography{IEEEabrv,refs}

\end{document}